# Development and Evaluation of the Institutionally-Farmed Research On-line Repository and Management System (InFORMs) towards Knowledge-Sharing and Utilization

**Billy S. Javier[1†], Leo P. Paliuanan, Corazon T. Talamayan, James Karl A. Agpalza, and Jesty S. Agoto**

*Cagayan State University, Aparri, Cagayan 3515 Philippines*

**Summary**
Knowledge derived from scientific discoveries and experiments are essential drivers to information management. This study was geared at promoting institutionally-farmed researches as a critical driver to utilization and re-production of new knowledge thru the use of the developed web application, Institutionally-Farmed Research Online Repository and Management System using the SCRUM development methodology. Assessing the application's compliance to ISO 25010 software quality characteristics, the study described the responses of consented participants after experiencing the use of the application in a certain duration. Key features that aids in the ease of use, access to, and management of the research resource emerged. From the responses, the developed application evidently suggests its usability and compliance to standards from the participants. The application was limited to showcasing all research resources produced in the University, in the last 10 years, after having gone through approval of the research review committee prior inclusion in the database. Maximizing the web application for knowledge-sharing and utilization is commended, advancing instruction and knowledge economy.

## 1. Introduction

New knowledge generated from research and development activities are useless information unless these scientific discoveries are made known to people and communities, adopt these technologies, and capacitate the countryside towards rural development. As data has become the new oil that drives the knowledge economy, it is vital that knowledge obtained from the scientific processes be transcended as common knowledge to communities. Information, education, and communication or IEC development could best thrive the sharing of research outputs, utilization of new generated technologies, and productivity-enhancement in the society. With the convergence of internet technology and the proliferation of social media and communication tools, research outputs should traverse the wider audience utilizing information and communications technologies, generating greater impact to the community.

The University mantra on producing publishable research outputs has led researchers and professionals to write scientific papers from conducted projects aligned to regional and national development goals. Publication as the major end result of research activities has been challenging university targets, SUC levelling, and indicators to normative financing. Hence, this project looks at maximizing outputs of research through knowledge sharing possible thru information, communications and education (IEC) initiatives. Generated knowledge from researches extended thru publication, online open access journals, and online libraries serves as important references and benchmark ideas to the conduct of further research projects. As instruction is always based from outputs or generated knowledge from research, this project would enable academicians, professors, and instructors to utilize research results as instructional material and as basis for the conduct of classroom-based researches.

Various research outputs have been generated by the faculty-researchers and students of Cagayan State University [1] through its years of existence. The multidisciplinary researches that has been produced has been either published or shelved for years. Some of these has been conveyed majority into various national and international conferences. Some others may have been cascaded to local meetings, seminars, or training-workshops. However, students as well as faculty members may not be well informed of how far CSU has been producing new knowledge from the classrooms. It may be known to very few faculty-researchers as part of the audience in in-house reviews, regional, national, and international conferences. The stakeholders of the University particularly the students are not widely disseminated of what were the generated new knowledge from these researches since majority of the outputs are not widely disseminated to them despite the fact that this knowledge came from the University experts and researchers.

The utilization and dissemination of these new knowledge may perish instead of being published and form part of the dissemination of new knowledge. Also, it is important to monitor and control the increasing gap between the information and the end-user that causes the difficulties of length, variety, velocity, veracity, its implementation and challenges [2]. New knowledge from these researches may have been limited to the shelves, to the journals, if published, and to the certain libraries or offices. This generated knowledge should have been available in various media capturing the interest of both students and faculty members to further advanced instruction, generation of research, and innovation of new technologies both benefitting the students, institution, and the nation in general. In fact, the utilization of these knowledge would bring a shift in the approaches to teaching and learning from cognitivism to socio-constructivism, connectivism and heutagogy [3].

Scientific publications are all around us, but accessibility and security concerns are problematic [4]. Merging the older tradition of compact, small publications with the Open Web Platform's ubiquitous usability, addressability, and interconnectedness is of great benefit. Online growth has impacted various aspects of our lives, such as communication, information sharing, social activities, and so on [5]. It has goals and it is becoming a need. As a type of knowledge management system, the web portal provides an excellent medium for information sharing and search [6].

Hence, to address the gap and to hasten the research culture from the classroom level to the University level, this project, Institutionally-Farmed Research Online Repository and Management System (InFORMs) is an educational IEC media that would aid instruction, research, and even extension. Venturing on the wide utilization of internet technologies, the project aimed at designing and development of an online knowledge sharing and management portal that maintains a database or online repository of research outputs produced from different disciplines within the University. It is hoped that the repository will advance instruction, generate new knowledge, and widen information dissemination practices along research outputs. Furthermore, developed IEC materials written by experts in the field of fisheries, education, information and communications technology, hospitality industry management, criminal justice and governance, business, entrepreneurship and accountancy, as well as industrial technology needed promotion thru multimedia technology development.

With the utility of the web and the internet, this project focuses on the development of the research or knowledge-sharing portal for the students and faculty of the University as it is a necessary tool for the students and faculty in the University in helping them create or publish researches accessible in the web and help others in accessing past researches for their references. In fact, web development becoming a global knowledge web development, strengthens various knowledge-sharing aactivities abd fosters broader global knowledge-sharing initiatives [1], especially in this disrupted COVID19 pandemic.

### 1.1 Objectives

Generally, the study aimed to design, develop, and implement an Institutionally-Farmed Online Research Repository and Management System (InFORMs) as a University means of an information, education, and communication (IEC) technologies geared towards dissemination of research outputs and utilization of newly generated technologies among communities. In addition, the extent of compliance to ISO 25010:2011 software quality have been made. More so, it is hoped that thru the project, there is an increase promotion of new knowledge and/or research outputs as well as research-based IEC materials thru multimedia technology development and make it available online.

## 2. Methodology

### 2.1 Research Design

The SCRUM development methodology was employed in the design and development of online repository involving the key processes of the system development life cycle, planning, analysis, design, development, and implementation. The descriptive design described the extent of compliance to ISO 25010:2011 [7] software quality characteristics of the developed InFORMs.

### 2.2 Respondents

Qualifying some inclusion criteria, there were 10 experts and/or practitioners in the IT industry evaluated the usability, acceptability, and satisfaction following the International Standards Organization [7]. Two (2) faculty-researchers with five students representing each College and the RDE staff formed part of the respondents evaluating the acceptability, usability, and satisfaction.

### 2.3. Instruments

The study utilized a structured questionnaire adapting the usability, acceptability, and satisfaction following the International Standards Organization (ISO/IEC) 25010:2011 software quality. In the design and development of the repository, the XAMPP development framework was employed utilizing open-source tools.

### 2.4. Statistical Tools

Frequency counts, weighted means and percentages were used to describe the data. T-test was used to test the differences of the assessment between the respondents.

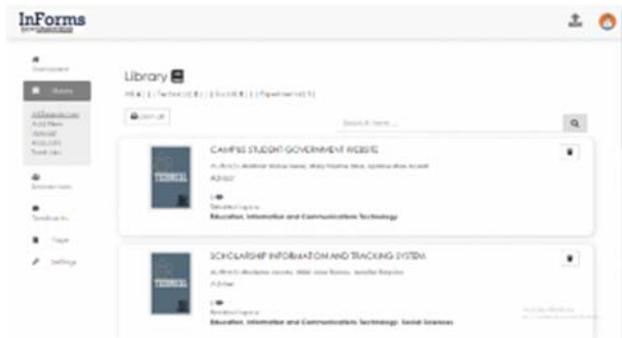

Fig 1 Dashboard of the InFORMS

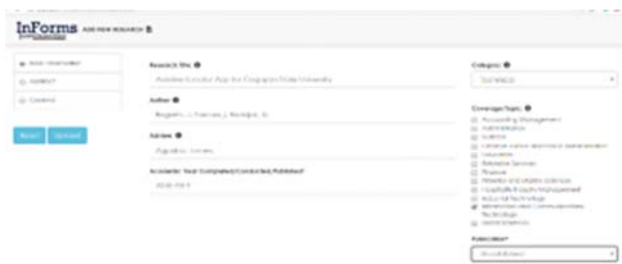

Fig 2 Adding, updating, and managing research resource

## 3. Results and Discussion

### 3.1. The Application – InFORMS: Institutionally-Farmed Research Online Repository and Management System

The web-based application, InFORMs, have been developed to provide students, researchers, and teachers an access to scientific experiments and articles that were documented in the University. The term "Farmed" was coined for institutionally produced or formulated researches or scientific undertaking which have undergone review by the Research Committee. The InFORMs required access to accounts by provision of security in adding and management of research resources. Users could view an abstract of all archives in the project allowing them to review the article prior full access. It provided an intuitive dashboard for users to ably maximize the portal with ease of managing their access and searching activities (Fig 1).

Each department that handles categories of the research resource allows addition and updating of articles and manuscripts in the archives (Fig 2). The addition of the new research resource passes thru the formed research committee prior its final inclusion in the InFORMs database (Fig 3). The department adding the new entry are notified of the processes.

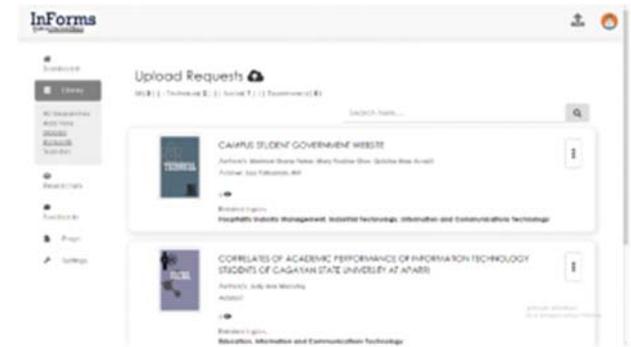

Fig. 3. Uploading for database inclusion

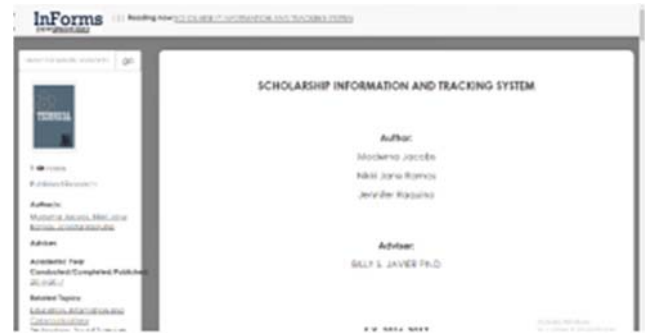

Fig. 4. Full view and print options

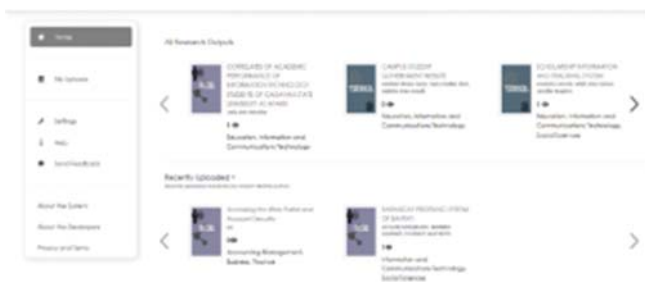

Fig 5. Informative categorization and classification

Full access of the research resource readily available in the archive requires the student, teacher, or requesting party to login with their account details. Only then can the user be able to view the full resource (Fig 4). Requesting party wanting to make copies of the research article from the archives once given the access would be able to download copy for the user. In addition, simple article information provided users of the number of reads or views in the resources and each categories (Fig 5, 6). This provided the administration to strategizing the use of the different research resources as tool to teaching and learning. The assistive feature includes retrieval of account, feedback form, and search and view. Moreover, relevant information about the University, Colleges, Terms, and Contacts are made available for all users. This applies clustering and classification algorithms for easy user interfacing and utilization.

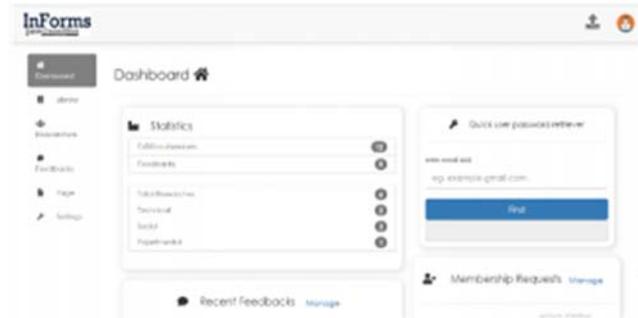

Fig 6 Classification and Clustering Info

### 3.2. The extent of compliance of the developed application system to ISO 25010:2011 software quality standards

The assessment on the extent of compliance of the web application, InFORMs, to ISO 25010 software quality characteristics have been conducted after a formal consent, presentation, and actual usage. The characteristics or dimensions in the evaluation of the analysis are defined in terms of functional sustainability, performance efficiency, compatibility, usability, reliability, security, maintainability, and portability. Separate results were presented for the IT Experts, for the RDE staff and Faculty Members, and for the students.

The IT Experts rated the functional suitability of the developed InFORMs to the high extent with a mean of 3.72, particularly underscoring its completeness, appropriateness, and correctness. Meanwhile, the users generally assessed the functional suitability aspects of the InFORMs compliant to a very high extent with a mean of 4.24, particularly along with completeness, correctness, and user access. This tends to imply that the proposed system conforms to the need for a complete, appropriate, and correct system for the students, thus generates the expected results by the participants. This finding agrees with the study of the team of Sharma, Maheshwari, Sengupta, & Shukla (2003) which highlighted the systems' compliance to functional completeness, correctness, and appropriateness to users as emphasized in a study [8]. Thus, satisfying functionality of the system aids to user usability.

With a mean of 3.78, the compliance of the system in terms of its reliability was found to be at the high extent particularly along with fault tolerance, maturity, and supportability. The finding suggests that the reliability of the system is important especially handling faults within, errors corrected over time, and recoverability. Reliability features of the system as assessed by the participants was found similar to the study [9] titled "Reliability of web-based information system in inquiry projects" which highlighted the reliability of systems affecting the usability of such. In fact, in the study "Applying the ISO Standard in Assessing the Quality of Software Systems" [10], were most participants felt that the software met the reliability of criterion, likely due to the knowledge and understanding of the users, conforms to the findings herein.

While IT experts generally assessed the system's compliance to usability requirement to the high extent with a mean of 3.82 based on the experiences, the users generally assessed the system's usability compliant to a very high extent with a mean of 4.37. The results highlighted the system's compliance to the high extent the operability, learnability, recognizability, and accessibility of the InFORMs. The results imply that the ability of the user to learn with ease, operate with minimal or no intervention, and easily understood by the wider audience has significant impacts on usability of the system, thus, is usable at the context of the users. The findings agree with the study [11] which underscored attractiveness, learnability, controllability, helpfulness, and efficiency meeting users' needs. Hence, the developed InFORMs has met the usability requirement very satisfactorily.

Table 1 Summary of the Assessment on the extent of compliance to ISO 25010

| Criteria | IT Experts (N=10) | | Users (N=30) | |
|---|---|---|---|---|
| | W. Mean | Description | W. Mean | Description |
| Functional Suitability | 3.72 | High Extent | 4.24 | Very High Extent |
| Reliability | 3.78 | High Extent | 4.40 | Very High Extent |
| Usability | 3.82 | High Extent | 4.37 | Very High Extent |
| Performance Efficiency | 3.87 | High Extent | 4.39 | Very High Extent |
| Maintainability | 3.92 | High Extent | 4.32 | Very High Extent |
| Portability | 3.90 | High Extent | 4.20 | High Extent |
| Security | 3.48 | High Extent | 4.36 | Very High Extent |
| Compatibility | 3.57 | High Extent | 4.33 | Very High Extent |
| Overall Weighted Mean | 3.76 | High Extent | 4.33 | Very High Extent |

The compliance of the developed InFORM in terms of performance efficiency has been assessed by the IT experts compliant to a high extent with a mean of 3.87. This criterion was greatly influenced by the capacity of the system to maximize the utilization of available resources efficiently. Results agree with the findings [10] ensuring quality in any product or service requires setting standards and ensuring that these are adhered to. These findings imply that users looked forward to utilizing in a timely manner and with ease while resources are maximized.

The system in terms of maintainability was found compliant to the high extent with a mean of 3.92. Analyzability, modifiability, and reusability were sub-criterion supporting this level of assessment made by the participant. The results tend to imply that the system has satisfied the need to be easily maintained without fear of the occurrence of faults, providing means to easily diagnose faults while functioning properly with ease of use and systematic processes. In fact, within the codes are code comments assisting the software maintainability and reliability, similar to Tan's (2015) claim that leveraging to improve software maintainability and reliability, code comments should contain rich information for users.

Similarly, the users assessed the portability criteria compliant to a high extent with a mean of 4.20, underscoring herein replaceability, and installability. This finding tends to suggest that the system has ably allowed a user to easily set-up the application, and ably conform to the standards in moving through another environment. It is also in agreement with the study "Key Factors for Developing a Successful E-Commerce Website" [12], emphasizing the ability of the system to be working on another environment attributes to the successful delivery of a website.

In terms of the security, the assessment highlighted integrity and authenticity an important feature greatly affecting the system's compliance to security. Meanwhile, the users generally assessed the security feature compliant to a very high extent with a mean of 4.36. The assessment also highlighted confidentiality, non-repudiation, and integrity of data or records. This tends to imply that the application has provided security and safety mechanisms attaining user trust. In fact, the findings agree with the results [13] underscoring security points and procedures at right on time phases of advancement and all throughout the product development cycle.

It is noteworthy to assert the system's compatibility having been found compliant to a high extent with a mean of 3.57 among the IT experts particularly highlighting the system's performance in a shared environment, as against 4.33 among users. It can be seen that a varying result of the assessment made is comparable. In particular, the users generally assessed the compliance of the InFORMs to a very high extent with a mean of 4.33 as compared to the IT Experts with a mean of 3.76. This is in agreement with [11] that posted that different perspectives derived from the area of specialization gives different evaluation, the assessment made herein evaluates to a standards, qualifying to a known norm of software quality worldwide, the ISO 25010 [14]. Normally, with the technical difference due to experience and/or level of skills, the web application has sustained a significant level of conformity to software quality standards. Similarly, the perceived compliance with the norm reflects the overall positive impression [15]. Further, the results agree with [10] that systems should not only conform to but extends beyond standards of the ISO 25010.

It can be seen at Table 1 that a varying result of the assessment made is comparable. In particular, the users generally assessed the compliance of the InFORMs to a very high extent with a mean of 4.33 as compared to the IT Experts with a mean of 3.76. This is in agreement with [11] in a study, "Assessing the Usability of University Websites From Users' Perspectives", that posted that different perspectives derived from the area of specialization gives different evaluation, the assessment made herein evaluates to a standards, qualifying to a known norm of software quality worldwide, the ISO 25010 [14]. Normally, with the technical difference due to experience and/or level of skills, the Project InFORMS has sustained a significant level of conformity to software quality standards. Similarly, the perceived compliance with the norm reflects the overall positive impression [15]. Further, the results agree with [10] that systems should not only conform to but extends beyond standards of the ISO 25010.

The findings suggest that the users highly recognize the potential of the InFORMs considering its potential for providing useful information to the students, enrichment of instruction, and providing inputs for research management purposes. It is believed that with the complementation viewed by the extent of compliance of the application by the users, the decision-support structure, in support to the findings of [16], is hoped to provide information and decision alternatives or actions by the top management. In addition, the use of web application as an IEC tool for knowledge-sharing agrees with literatures, to assist with the dissemination of data allowing interested parties access to view, explore, and utilized the outputs is both beneficial to the students, the management of the University, and the community.

Table 2 Differences in the assessment between groups

| Characteristics | Group | Mean | t | p-value | Remarks |
|---|---|---|---|---|---|
| Functional Suitability | IT Expert | 3.72 | -4.596 | 0.000 | Significant |
|  | User | 4.24 |  |  |  |
| Reliability | IT Expert | 3.78 | -5.262 | 0.000 | Significant |
|  | User | 4.40 |  |  |  |
| Usability | IT Expert | 3.82 | -4.522 | 0.000 | Significant |
|  | User | 4.37 |  |  |  |
| Performance Efficiency | IT Expert | 3.87 | -2.844 | 0.007 | Significant |
|  | User | 4.39 |  |  |  |
| Maintainability | IT Expert | 3.92 | -2.779 | 0.008 | Significant |
|  | User | 4.32 |  |  |  |
| Security | IT Expert | 3.90 | -2.401 | 0.021 | Significant |
|  | User | 4.20 |  |  |  |
| Portability | IT Expert | 3.48 | -5.290 | 0.000 | Significant |
|  | User | 4.36 |  |  |  |
| Compatibility | IT Expert | 3.57 | -3.722 | 0.003 | Significant |
|  | User | 4.33 |  |  |  |

The mean difference, as seen in Table 2, in the assessment on the extent of compliance to ISO 25010 suggests a varying assessment approval from the users than that of the IT experts. From the table, the p-values for each criterion is less than 0.05 or 0.01, implying that the difference in means is statistically significant at the 95 percent or 99 percent level of significance. This finding tends to suggest that the developed InFORMs ably provided the user a considerable understanding of the system with the provision of the features that were usable and supportive to the conduct of the activities or processes in monitoring and data management of the aramang production, catch, and yield. Meanwhile, with the technical exposure in hand, the assessment of the IT experts tends to suggest the level of technical features expected in the InFORMs thereby affecting their assessment far different with the users.

## 4. Conclusion and Implications

Providing enough support available for the students and faculty members in searching/accessing of scientific articles or experiments for references and in monitoring and managing of the research, the new system (InFORMs available for online use) will be one systematic solution that could aid students and faculty in terms of searching, finding specific study/research according to their interest. The assessment generally revealed the systems' compliance to ISO 25010:2011 software quality characteristics. The noble web application is one form to publish for knowledge-sharing rather being perished or wasted.

Knowledge derived from extensive research engagement from faculty members, to students and external collaborators are only made known if proper channel is made. Hence, considering the greater benefit of the project, it recommended to maximize the use of the project to make the service to students and faculty easier, reliable, faster, better and efficient to use with no cost especially utilized for instruction, research and development. Eventually, exploiting the system will definitely allow users to explore new discoveries, uncover new knowledge based on the gaps identified from current research outputs, and be able to streamline knowledge-sharing through the InFORMs. In the greater side, the data as the new oil emanates from the repository, thereby impacting academic institutions for instructional purposes, and translating recommendations to policies improving the lives of people and benefiting the communities.

## Acknowledgement

The authors are very grateful to the Cagayan State University administration for the kind support and grant. Special intention is also extended to the RDE Office for the technical assistance, the group of reviewer, experts, students, library staff, and other support group for the participation and assistance.

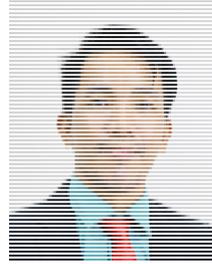

**Billy S. Javier** received the B.S. and Master's degrees in Information Technology, from Cagayan State Univ. in 2006 and 2008, respectively, as well as the PhD in Education major in Educational Management in 2013, and Doctor in Information Technology as CHED K12 SGS Grantee in 2019 at SPUP. He served various positions in the Univ as Dean of the CICS, Coordinators for Research, KTM, and Internationalization. His research interest includes IT and its application to education, fisheries, social sciences and governance. He is an associate member of NRCP, PSITE, PhilAAST, and other scientific and professional organizations in the Philippines. He is currently holding Assoc Professor Rank at Cagayan State University at Aparri.

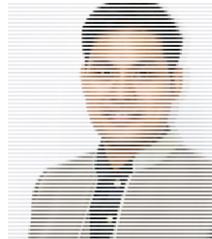

**Leo P. Paliuanan** is currently an instructor in web development and databases at Cagayan State University Aparri. He received his BS and MS in Information Technology in 2009 and 2013 respectively. He is a full stack developer engaged in managing web sites, information management and databases.

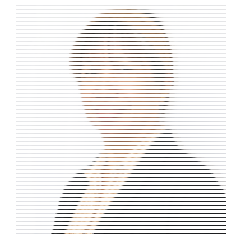

**Corazon T. Talamayana** is holds a degree in B.Sc, Computer Science, Masters in IT, and PhD in Educational Management. She served as Dean of the College of Information and Computing Sciences. She is a member of the NRCP and PSITE. Her research interest are in educational management, applications of IT in education, and mathematics.

**James Karl A. Agpalza** received his BS and MS in Information Technology at Cagayan State University at Aparri in 2009 and 2013 respectively. His works are in web design and graphics design. He is a member of the PSITE – a national organization of IT professionals.

**Jesty S. Agoto** received his BS and MS in Information Technology in 2010 and 2014 respectively at Cagayan State University at Aparri. His